\documentclass{book}

\usepackage[paperwidth=17cm, paperheight=24cm, includeheadfoot,
            left=2cm, right=2cm, top=2cm, bottom=1.3cm, twoside]{geometry}

\usepackage%
 {
  amsmath, amssymb, appendix, cite, enumitem, fancyhdr, fancyvrb,
  graphicx, import, lineno, longtable, remreset, shorttoc, url
 }

\usepackage[rightcaption]{sidecap}
\usepackage{caption}
\usepackage{algpseudocode}

\newcommand{\tmpstring}{}
\newcommand{\settmpstring}[1]{\renewcommand{\tmpstring}{#1}}
\newcommand{\SourceCodeLines}[1]
 {%
  \settmpstring{{\ttfamily\bfseries\tiny\theFancyVerbLine}}
  \ifnum#1>9
    \settmpstring
     {\parbox[b]{7.5pt}{\ttfamily\bfseries\tiny\rightline\theFancyVerbLine}}
  \fi
  \ifnum#1>99
    \settmpstring
     {\parbox[b]{11.2pt}{\ttfamily\bfseries\tiny\rightline\theFancyVerbLine}}
  \fi
 }

\def\thepart{\Alph{part}}

\makeatletter
\@removefromreset{figure}{chapter}
\makeatother

\makeatletter
\renewcommand{\thefigure}{\@arabic\c@figure}
\makeatother

\makeatletter
\@removefromreset{table}{chapter}
\makeatother

\makeatletter
\renewcommand{\thetable}{\@arabic\c@table}
\makeatother

\makeatletter
\@removefromreset{equation}{chapter}
\makeatother

\makeatletter
\renewcommand{\theequation}{\@arabic\c@equation}
\makeatother

\usepackage{tocloft} % packages should be loaded after redefinition
\usepackage{minitoc} % of sectioning and other formating commands

\cftsetpnumwidth{1.73em}

\makeatletter
\renewcommand{\@tocrmarg}{4em}
\makeatother

\mtcsetformat{minitoc}{tocrightmargin}{2.55em plus 1fil}

\newcommand{\Author}{}
\newcommand{\AuthorLastName}[1]{\renewcommand{\Author}{#1}}

\def\Title#1{\chapter[\thepart\thelecture\ \protect\mbox{\protect%
\parbox[t]{110mm}{#1 \textit{(\Author)}}}\smallskip]{\Large #1}}

\newsavebox{\shortTitleBox}
\def\shortTitle#1{\savebox{\shortTitleBox}{#1 \textit{(\Author)}}}

\newcounter{lecture}[part]

\newcommand{\SwitchToFancy}
 {%
  \pagestyle{fancy}
  \fancyhf{}
  \fancyhead[OR]{\rightmark}
  \renewcommand{\headrulewidth}{0.4pt}
  \fancyfoot[OR]{\thepage}
  \fancyfoot[EL]{\thepage}
  \fancyhead[EL]{\usebox{\shortTitleBox}}
 }

\makeatletter
\def\thebibliography#1
{
 \section*{References\@mkboth{References}{References}}
 \list{[\arabic{enumi}]}
  {
   \settowidth\labelwidth{[#1]}\leftmargin\labelwidth\advance\leftmargin
   \labelsep\usecounter{enumi}
  }
 \def\newblock{\hskip .11em plus .33em minus .07em}
 \sloppy\clubpenalty4000\widowpenalty4000\sfcode`\.=1000\relax
}
\makeatother

\makeatletter
\renewcommand{\@makefntext}[1]{\setlength{\parindent}{0pt}%
\begin{list}{}{\setlength{\labelwidth}{1.5em}%
\setlength{\leftmargin}{\labelwidth}%
\setlength{\labelsep}{3pt}\setlength{\itemsep}{0pt }%
\setlength{\parsep}{0pt}\setlength{\topsep}{0pt}%
\footnotesize}\item[\hfill\@makefnmark]#1%
\end{list}}
\makeatother

\input{syntaxHighlighting_melchert.tex}

\begin{document}

\dominitoc

\faketableofcontents

%%% Correction of an unfortunate interaction of tocleft and minitoc packages
%%% (section entries in minitocs in bold)
\renewcommand{\cftsecfont}{\bfseries}
\renewcommand{\cftsecleader}{\bfseries\cftdotfill{\cftdotsep}}
\renewcommand{\cftsecpagefont}{\bfseries}

%%% Page style for first page of Template
\fancypagestyle{plain}
 {
  \renewcommand{\headrulewidth}{0pt}
  \fancyhead[OL]
   {
    \vspace{-8.5pt}
    \textit{Lecture given at the International Summer School}\\
    \textit{\emph{Modern Computational Science 2012 -- Optimization}}\\
    \textit{University of Oldenburg, Germany, August 20-31, 2012}
   }
 }

\stepcounter{lecture}
\setcounter{figure}{0}
\setcounter{equation}{0}
\setcounter{table}{0}

%% AUTHOR DETAILS {{{1
%%%%%%%%%%%%%%%%%%%%%%%%%%%%%%% Author and Title %%%%%%%%%%%%%%%%%%%%%%%%%%%%%%
\AuthorLastName{Melchert}
\Title{Minimum weight spanning trees of weighted scale free networks}
\shortTitle{Minimum weight spanning trees}
\SwitchToFancy
\bigskip
\bigskip
%%%%%%%%%%%%%%%%%%%%%%%%%%%%%%%% Author Address %%%%%%%%%%%%%%%%%%%%%%%%%%%%%%%
\begin{raggedright}
  \itshape Oliver Melchert\\
  Institute of Physics\\
  Faculty of Mathematics and Science\\
  Carl von Ossietzky Universit\"at Oldenburg\\
  D-26111 Oldenburg\\
  Germany
  \bigskip
  \bigskip
\end{raggedright}
%%}}}1

%% ABSTRACT {{{1 
\paragraph{Abstract.}
In this lecture we will consider the minimum weight spanning tree (MST) problem, 
i.e., one of the simplest and most vital combinatorial optimization problems.
We will discuss a particular greedy algorithm that allows to compute a MST for
undirected weighted graphs, namely \emph{Kruskal's} algorithm, and we will study the
structure of MSTs obtained for weighted scale free random graphs. This is meant 
to clarify whether the structure of MSTs is sensitive to correlations between 
edge weights and topology of the underlying scale free graphs.
\medskip

\begin{center}
  \framebox
   {%
    \parbox{0.9\textwidth}
     {%
      The lecture is supplemented by a set of {\tt Python} scripts that allow 
      you to reproduce figures 4 and 5 shown in the course of the lecture.
      The supplementary material can be obtained from 
      the MCS homepage (see Ref.\ \cite{supMat}).%
     }%
   }
\end{center}

%\newpage
%%}}}1

\minitoc

\section{Recap: Nodes, Edges, and Graphs}

\paragraph{Nodes:}{ 
A \emph{node set} $V$ is a collection of elements $i$, termed \emph{nodes} 
(also called \emph{sites} or \emph{vertices}). The number of nodes in a node set
is subsequently referred to as $N\!=\!|V|$.}

\paragraph{Edges:}{
An \emph{edge} (or \emph{arc}) $e_{ij}$ consists of a pair of nodes $i,j\!\in\!V$. 
Edges can either be directed or undirected. In the former case the corresponding
node pair is ordered, i.e., $e_{ij}\!=\!(i,j)$ with $(i,j)\!\neq\!(j,i)$, while in the latter 
case the node pair is unordered, i.e., $e_{ij}\!=\!\{i,j\}$ with $\{i,j\}\!\equiv\!\{j,i\}$.
In the following, if not stated otherwise, the term edge will always refer to 
an \emph{undirected edge}, i.e., $e_{ij}\!=\!\{i,j\}$.
The edge $e_{ij}$ is said to be \emph{incident} with the nodes $i$ and $j$ and
it \emph{joins} both nodes.
Two distinct nodes are said to be \emph{adjacent}, 
if they are incident with the same edge. Similarly, two distinct edges are adjacent, 
if they have a node in common.

Consequently, an \emph{edge set} $E$ is a collection of elements $e_{ij}$ with 
$i,j\!\in\!V$. 
In what follows, $E$ is not allowed to contain \emph{self-edges}, i.e., edges 
of type $e_{ii}$, 
or multiple parallel edges. 
The number of edges in an edge set is referred to as $M\!=\!|E|$.}
%
%%%%%%%%%%%%%%%%%%%%%%%%%%%%%%%%%%%%%%%%%%%%%%%%%
%% FIGURE:	example graphs
%%		- undirected graph
%%		- directed graph
%%
%%		\label{fig:graphs1_ab}
%%%%%%%%%%%%%%%%%%%%%%%%%%%%%%%%%%%%%%%%%%%%%%%%%
%{{{1
\begin{SCfigure}[][b]
%  \centering
  \includegraphics[width=0.49\textwidth]{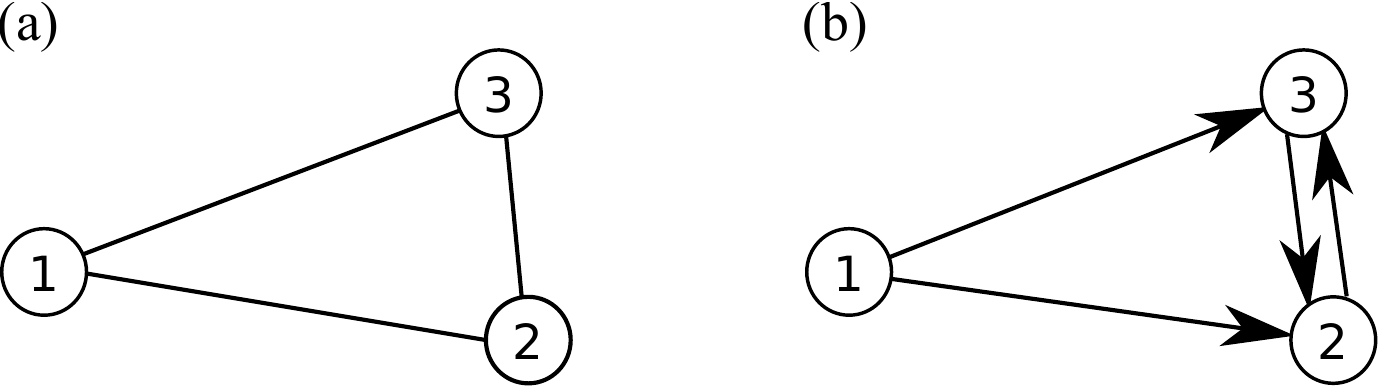}
\caption{\footnotesize{Directed and undirected example graphs. 
(a) undirected graph with node set $V\!=\!\{1,2,3\}$ and edge set $E\!=\!\{\{1,2\},\{1,3\},\{2,3\}\}$, 
(b) directed graph with node set $V\!=\!\{1,2,3\}$ and edge set $E\!=\!\{(1,2),(1,3),(2,3),(3,2)\}$.}}
\label{fig:graphs1_ab}
\end{SCfigure}
%}}}1%%%%%%%%%%%%%%%%%%%%%%%%%%%%%%%%%%%%%%%%%%%%

\paragraph{Graphs:}{
A \emph{graph} $G\!=\!(V,E)$ is a tuple that consists of a node set $V$ 
and an edge set $E$. Depending on the characteristics of the edge set, 
a graph can either be directed or undirected (see Fig.\ \ref{fig:graphs1_ab}(a),(b)). 
In the following, if not stated otherwise, the term graph will always refer 
to an \emph{undirected graph}.

Within a graph, a particular node can have several adjacent nodes, given by 
the set ${\rm adj}(i)\!=\!\{j~|~e_{ij}\!\in\!E \}$. In this context, the \emph{degree} 
$d(i)\!=\!|{\rm adj}(i)|$ of node $i$ measures the number of its neighbors. 
Regarding directed graphs, one might
distinguish between the \emph{in-} and \emph{out-}degree of a node, e.g.\ node $3$ 
of the directed graph in Fig.\ \ref{fig:graphs1_ab}(b)  has $d_{\rm out}(3)\!=\!1$ and 
$d_{\rm in}(3)\!=\!2$ while the corresponding node in the undirected graph 
(Fig.\ \ref{fig:graphs1_ab}(a)) simply has $d(3)\!=\!2$.

Further, \emph{mappings} can be used to relate additional information to the graph. 
Therefore, nodes as well as edges can be addressed. E.g., a \emph{weight function} 
$\omega\!:E\rightarrow \mathbb{R}$ can be used in order to assign a certain 
weight $\omega_{ij}\!\equiv\!\omega(e_{ij})$ to each edge $e_{ij}\!\in\!E$, see Fig.\ \ref{fig:MST_smallExample}. 
Such a weight might be interpreted as distance between the two nodes or as a cost to get 
from one node to the other. The triple $G\!=\!(V,E,\omega)$ is then called a 
\emph{weighted graph}.

A \emph{subgraph} $G_{\rm sub}\!=\!(V_{\rm sub},E_{\rm sub})$ is obtained from a graph 
$G$ by deleting a (possibly empty) subset of its nodes and edges. That means, for
a subgraph it holds that $V_{\rm sub}\!\subseteq\!V$ and $E_{\rm sub}\!\subseteq\!E$.

A \emph{cut} $(C,V\setminus C)$ on an undirected graph $G=(V,E)$ is a partition of its
nodeset. An edge $\{i,j\}\in E$ \emph{crosses} the cut if $i \in C$ and $j \in V\setminus C$ (or vice versa).
A cut \emph{respects} a subset $E_{\rm sub} \subseteq E$ if there is no edge in $E_{\rm sub}$ that 
crosses the cut.
An edge $\{i,j\}$ is called \emph{candidate} regarding the cut $(C,V\setminus C)$ if it crosses
the cut and has minimum weight amongst all edges that cross the cut.
As an example consider the cut $(C=\{0,1,2\},V\setminus C = \{3\})$ (the nodes in the set $C$ 
are colored in grey) in Fig.\ \ref{fig:MST_smallExample}(a). 
Therein, the edges $\{1,3\}$ and $\{2,3\}$ cross the cut. Amongst those two, $\{2,3\}$ signifies the candidate edge
since it has the smaller weight.}

\section{Minimum weight spanning trees (MSTs)}

Given an undirected, connected and weighted graph $G=(V,E,\omega)$ 
where $N=|V|$ and $M=|E|$ signify the number of nodes and edges of $G$, respectively, 
and where $\omega$ assigns a weight to each edge. 
Compute a ``minimum weight spanning tree'' $T$ \cite{kruskal1956,clrs2001,ahuja1993}.

\begin{SCfigure}[][b]
%  \centering
  \includegraphics[width=0.5\textwidth]{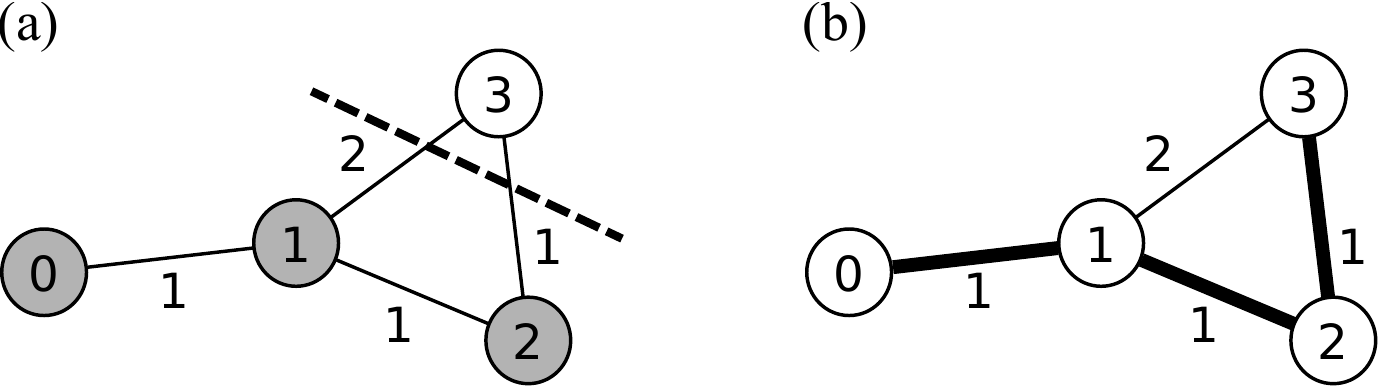}
  \caption{ 
(a) The dashed line illustrates the cut $(\{0,1,2\},\{3\})$ (for more details, see text). 
(b) example of a MST (bold black lines) for a small undirected weighted graph. 
        \label{fig:MST_smallExample}
          }
\end{SCfigure}

\paragraph{Minimum weight spanning tree (MST):} A MST $T$ is a connected, loopless subgraph of $G$,
 consisting of $(N-1)$ edges, connecting all $N$ nodes, thereby minimizing the sum of the
edge weights. 
Note that there are several ways to represent a MST. Here, we will represent it by means of 
the set $T$ of edges from which it is build up.
For the small undirected graph shown in Fig.\ \ref{fig:MST_smallExample}(b), the MST (indicated by bold black lines) 
reads $T=\{\{0,1\}, \{1,2\}, \{2,3\}\}$ and its corresponding weight is 
$\omega_T= \sum_{\{i,j\}\in T} \omega_{ij} = 3$.
Depending on the precise topology of $G$ and the distribution of edgeweights, $T$ is not necessarily unique.

\paragraph{Relevance of the MST problem:}
The minimum weight spanning tree (MST) problem is one of the simplest and most 
vital combinatorial optimization problems. As such, it arises in a vast number of
applications. Either as a ``standalone'' problem or as subtask of a more intricate
problem.
The generic problem that is solved by an MST algorithm reads:
``Connect a set of points using a minimum-weight set of edges''.
As such, broadcasting problems on networks typically relate to finding the MST on an 
appropriate weighted graph. E.g., consider a network where the edgeweights signify time
delays (or transmission costs) between the respective incident nodes. Then, 
an ``efficient'' or ``optimal'' broadcasting of a message is made sure if the message is
transmitted along the MST edges. The latter connects all nodes by using as few edges as 
possible, thereby minimizing the total time delay (or transmission cost).
Further, MSTs find application in the context of single linkage cluster analyses \cite{gower1969}.

\section{Computing MSTs via a ``greedy'' strategy}

\paragraph{Greedy problem solving strategy:} 
At any time (during the solution procedure for a given problem) where a decision is required, 
a \emph{greedy} problem solving strategy makes the locally optimal choice.
Further, unlike in \emph{backtracking} approaches, once a decision is made it is not 
revised later on.
Albeit such an approach might fail to find a globally optimal solution for your particular
optimization problem, an optimal solution
for the MST problem can very well be obtained via efficient greedy algorithms.
A generic greedy approach to compute a MST for a given network $G$ reads:

\begin{algorithmic}[1]
\Statex {\bf algorithm} MST\_greedy($G$)
\vspace{0.3cm}
\State $T=\{\}$
\While{ $T$ is not a MST}
  \State pick feasible edge $e \in E\setminus T$ 
  \State $T\gets T\cup e$
\EndWhile
\State return MST $T$
\end{algorithmic}
\vskip 0.2cm
\paragraph{Cut-property:} In the above pseudocode the tricky part is to find proper selection criteria
for the edges that are added to the MST.
In this regard, a feasible edge has to satisfy the \emph{cut property}: Consider a weighted undirected
graph $G=(V,E,\omega)$. Let $T^\prime \subseteq E$ so that there is a MST $T\supset T^\prime$. 
Further, let $(C,V\setminus C)$ be an arbitrary cut that respects $T^\prime$ and let $\{i,j\}$
be a candidate regarding the cut. Then $\{i,j\}$ is a feasible edge that can be used in order 
to extend $T^\prime$. E.g., regarding the cut illustrated in Fig.\ \ref{fig:MST_smallExample}(a),
the edge $\{2,3\}$ is a feasible edge that can be used to amend (and also complete) the MST.

%\newpage
\section{Kruskal's algorithm}

\paragraph{Idea behind the algorithm:} The algorithm due to Kruskal \cite{kruskal1956} is a particular greedy algorithm that,
for an instance of an undirected, connected and weighted graph, obtains a MST.
The subsequent three steps summarize the solution strategy of Kruskal's algorithm:
\begin{description}
\item[Step 1:] In the beginning, each node forms a tree. Hence, there is a forest of $N$ single-node trees.
Further, an empty list $T$ is initialized that will keep all those edges that are part of the MST.
\item[Step 2:] Sort the edges in order of increasing weight. Visit the edges in order of increasing weight
and apply the following edge-selection criterion:
\item[Step 3:] If the currently considered edge $\{i,j\}$ has both endnodes in 
different trees, add the edge to $T$ and merge the respective trees. If $\{i,j\}$ has both endnodes
in the same tree, discard the edge. Proceed to the next edge or stop when there are no more edges to process.
\end{description}
After $M$ iterations of step 3 (i.e., as soon as all edges are processed), $T$ comprises the edges 
that comprise the MST.

\paragraph{Efficient implementation:}
Albeit the pseudocode and the description above seem to be very simple, an efficient implementation of 
Kruskal's algorithm is more tricky than it appears at first sight. 
In this regard, step 2 can be accomplished by using merge-sort yielding a sorted edge list in 
time $O(M \log M)$.
Further, in step 1 single-node trees (one tree for each node in the graph) need to be initialized, 
and in step 3 it must be possible to efficiently determine the tree to which 
a node belongs. Finally, two selected trees need to be merged to a single tree, occasionally.
These latter three tasks can be handled by using a \emph{union-find} data structure that 
features the following three operations:
\begin{description}
\item[Operation 1:] {\tt make\_set(i)}: Generates a tree consisting of the node $i$, only
\item[Operation 2:] {\tt find(i)}: Yields the ``name'' of the tree to which node $i$ belongs
\item[Operation 3:] {\tt union(a,b)}: Merges trees with names $a$ and $b$ to a new tree with
name $a$
\end{description}
For a connected graph,
the total time for the union-find operations (i.e., $N$ {\tt make\_set} operations and 
$O(M\!)$ {\tt find} and {\tt union} operations) can be approximated by
$O(M\!\log M\!)$
(note that this also depends on the precise implementation of the union-find data 
structure \cite{clrs2001}).
The full (worst case) running time of Kruskal's algorithm amounts to $O(M \log N)$. It summarizes the time
spent to sort the edges, perform $M$ find and $N-1$ union operations. 
A {\tt python} \cite{pyRef} implementation of Kruskal's algorithm is contained in the supplementary 
material \cite{supMat}.
Generically, {\tt python} uses \emph{Timsort} \cite{timsort}, a hybrid sorting 
algorithm based on \emph{merge sort} and \emph{insertion sort} \cite{clrs2001}.
Further, the implementation of the union-find data structure contained in 
the supplementary material uses \emph{union-by-size}: if two trees are merged by means
of a call to {\tt union}, the smaller tree (in terms of the number of nodes contained in the tree)
is added to the larger tree.
A particular implementation might read (see \cite{clrs2001}):
\begin{Verbatim}[fontsize=\small,fontfamily=txtt,commandchars=\\\{\}]
\PY{k}{def} \PY{n+nf}{mstKruskal}\PY{p}{(}\PY{n}{G}\PY{p}{)}\PY{p}{:}
	\PY{l+s+sd}{"""Kruskals minimum spanning tree algorithm }
\PY{l+s+sd}{        }
\PY{l+s+sd}{        algorithm for computing a minimum spanning }
\PY{l+s+sd}{        tree (MST) T=(V,E') for a connected, undirected }
\PY{l+s+sd}{        and weighted graph G=(V,E,w) as explained in }
\PY{l+s+sd}{        'Introduction to Algorithms', }
\PY{l+s+sd}{        Cormen, Leiserson, Rivest, Stein, }
\PY{l+s+sd}{        Chapter 23.2 on 'The algorithms of Kruskal and Prim'}
\PY{l+s+sd}{	}
\PY{l+s+sd}{	Input:}
\PY{l+s+sd}{	G       - weighted graph data structure}

\PY{l+s+sd}{	Returns: (T,wgt)}
\PY{l+s+sd}{	T       - minimum spanning tree stored as edge list}
\PY{l+s+sd}{        wgt     - weight of minimum weight spanning tree}
\PY{l+s+sd}{	"""}
	\PY{n}{uf} \PY{o}{=} \PY{n}{unionFind\PYZus{}cls}\PY{p}{(}\PY{p}{)}    \PY{c}{\PYZsh{} union find data structure}
	\PY{n}{T}\PY{o}{=}\PY{p}{[}\PY{p}{]}                    \PY{c}{\PYZsh{} list to store MST edges}

        \PY{c}{\PYZsh{} list of edges sorted in order of increasing weight}
        \PY{n}{K} \PY{o}{=} \PY{n+nb}{sorted}\PY{p}{(}\PY{n}{G}\PY{o}{.}\PY{n}{E}\PY{p}{,}\PY{n+nb}{cmp}\PY{o}{=}\PY{k}{lambda} \PY{n}{e1}\PY{p}{,}\PY{n}{e2}\PY{p}{:} \PY{n+nb}{cmp}\PY{p}{(}\PY{n}{G}\PY{o}{.}\PY{n}{wgt}\PY{p}{(}\PY{n}{e1}\PY{p}{)}\PY{p}{,}\PY{n}{G}\PY{o}{.}\PY{n}{wgt}\PY{p}{(}\PY{n}{e2}\PY{p}{)}\PY{p}{)}\PY{p}{)}

        \PY{c}{\PYZsh{} initialize forrest of sinlge-node trees}
        \PY{k}{for} \PY{n}{i} \PY{o+ow}{in} \PY{n}{G}\PY{o}{.}\PY{n}{V}\PY{p}{:}
	        \PY{n}{uf}\PY{o}{.}\PY{n}{makeSet}\PY{p}{(}\PY{n}{i}\PY{p}{)}

        \PY{c}{\PYZsh{} construct MST}
	\PY{k}{for} \PY{p}{(}\PY{n}{v}\PY{p}{,}\PY{n}{w}\PY{p}{)} \PY{o+ow}{in} \PY{n}{K}\PY{p}{:}
		\PY{k}{if} \PY{n}{uf}\PY{o}{.}\PY{n}{find}\PY{p}{(}\PY{n}{v}\PY{p}{)}\PY{o}{!=}\PY{n}{uf}\PY{o}{.}\PY{n}{find}\PY{p}{(}\PY{n}{w}\PY{p}{)}\PY{p}{:}
			\PY{n}{uf}\PY{o}{.}\PY{n}{union}\PY{p}{(}\PY{n}{uf}\PY{o}{.}\PY{n}{find}\PY{p}{(}\PY{n}{v}\PY{p}{)}\PY{p}{,}\PY{n}{uf}\PY{o}{.}\PY{n}{find}\PY{p}{(}\PY{n}{w}\PY{p}{)}\PY{p}{)}
			\PY{n}{T}\PY{o}{.}\PY{n}{append}\PY{p}{(}\PY{p}{(}\PY{n}{v}\PY{p}{,}\PY{n}{w}\PY{p}{)}\PY{p}{)}
                     
        \PY{k}{return} \PY{n}{T}\PY{p}{,} \PY{n+nb}{sum}\PY{p}{(}\PY{n+nb}{map}\PY{p}{(}\PY{k}{lambda} \PY{n}{e}\PY{p}{:} \PY{n}{G}\PY{o}{.}\PY{n}{wgt}\PY{p}{(}\PY{n}{e}\PY{p}{)}\PY{p}{,}\PY{n}{T}\PY{p}{)}\PY{p}{)}
\end{Verbatim}

%\newpage

\paragraph{Example graph:}

\begin{figure}[tb!]
%\begin{SCfigure}[][b]
\centerline{\includegraphics[width=1.\linewidth]{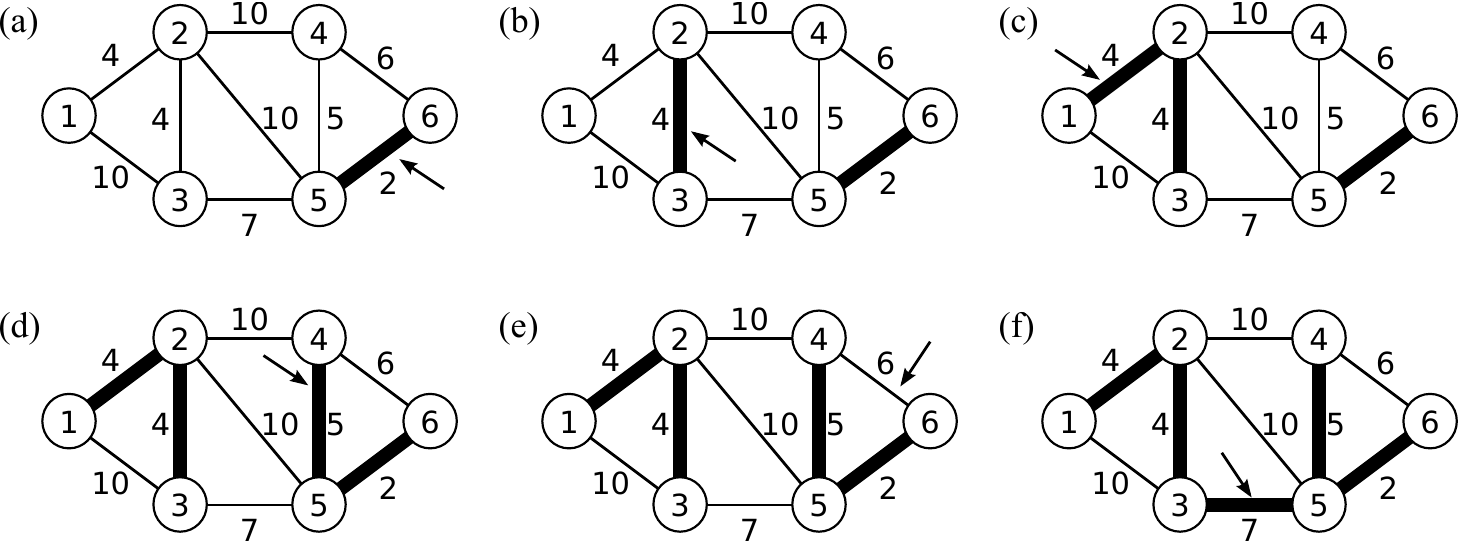}} 
\caption{Exemplary application of Kruskal's MST algorithm on a small example graph $G$
consisting of $6$ nodes and $9$ edges. The bold black edges belong to the forest that 
is grown in order to construct an MST of $G$. In the course of the algorithm each edge is
considered once (within the subfigures (a--f), a small arrow indicates the edge currently
considered). For each edge it is decided whether it can be used to extend the forest
constructed so far in order to yield an MST. For more details on the steps (a) through
(f), see text.}
\label{fig:alg_af}
%\end{SCfigure}
\end{figure}
In order to illustrate Kruskal's MST algorithm, consider the small graph $G=(V,E,\omega)$ 
consisting of $N=6$ nodes and $M=9$ edges, shown in Fig.\ \ref{fig:alg_af}(a--f).
In a first step, the graph edges are sorted in order of increasing weight, i.e.,
the set $K$ reads:
\begin{eqnarray}
K = \{   \{5, 6\}, \{2, 3\}, \{1, 2\}, \{4, 5\}, \{4, 6\}, \{3, 5\}, \{1, 3\}, \{2, 5\}, \{2, 4\}   \} \notag
\end{eqnarray}
Then, as a second step, the function {\tt make\_set} is used to initialize a forest of trees. In
the very beginning, each tree consists of a single node, only.
In the third step, the edge selection procedure is carried out until all edges are processed.
In detail, the following steps (illustrated in Fig.\ \ref{fig:alg_af}(a--f)) are carried out:
\begin{description}
\item[Step (a):] Edge $\{5,6\}$ is considered. It can be used to merge two distinct trees in the 
forest and is hence added to $T$.
\item[Step (b):] Edge $\{2,3\}$ is considered. It can be used to merge two distinct trees in the 
forest and is hence added to $T$.
\item[Step (c):] Edge $\{1,2\}$ is considered. It can be used to merge two distinct trees in the 
forest and is hence added to $T$. Note that $\{2,3\}$ and $\{1,2\}$ had the same weight. However,
no matter which of the two edges is picked first does not alter the structure of the final MST.
\item[Step (d):] Edge $\{4,5\}$ is considered and added to $T$. 
\item[Step (e):] Edge $\{4,6\}$ is considered. It does not connect two distinct trees. 
Adding it to the forest would introduce a cycle. Hence, the edge is not added to $T$ and 
the algorithm proceeds to the next edge in $K$.
\item[Step (f):] Edge $\{3,5\}$ is considered and added to $T$. 
\end{description}
All remaining edges would also lead to a cycle (as in step (e) above) and are hence not added to 
$T$.
Finally, after the algorithm terminates, the edgeset $T$ comprises a MST with weight $\omega_T=22$.

\section{MSTs of weighted scale free networks}

The subsequent simulations, reported in Ref.\ \cite{mcdonald2005}, were carried out to 
clarify whether the structure of MSTs change as a function of the correlations
between edge weights and network topology. 
(Here, as an exercise, I re-implemented the model studied in Ref.\ \cite{mcdonald2005} 
and re-performed their simulations and analysis. The resulting code is available as
supplementary material at \cite{supMat}. The figures presented below summarize the
results of the simulations performed via the code in the supplementary material.)

\paragraph{Simulation setup:}
For the numerical experiments scale free (SF) random networks were considered. 
These networks (also referred to as heterogeneous random graphs) are characterized 
by a power-law degree distribution $p_k \sim k^{-\gamma}$. The construction of
such SF networks via preferential attachment \cite{batagelj2005} yields an exponent $\gamma=3$.
In particular, SF networks containing $N=10000$ nodes were used (the 
number of edges connecting a newly created node to existing nodes during the 
preferential attachment procedure was set to $m=2$).

Further, two qualitatively different weight distributions were considered 
(note that the authors of Ref.\ \cite{mcdonald2005} considered more weight distributions,
however, the two weight distributions considered here suffice in order to answer the
question that kicked off the study):
\begin{description}
\item[Disorder type 1:] 
\begin{eqnarray}
\omega_{ij}=k_i k_j, \label{eq:wgtDistrib1}
\end{eqnarray}
where the weight associated to edge $\{i,j\}$ is
large if the degrees of its endnodes tend to be large, and,
\item[Disorder type 2:] 
\begin{eqnarray}
\omega_{ij}=1/k_i k_j, \label{eq:wgtDistrib2} 
\end{eqnarray}
where the weight related to an edge $\{i,j\}$ 
is large if the degrees of its endnodes tend to be small. 
\end{description}
For SF networks respecting the above two weight distributions, MSTs
are computed using Kruskal's algorithm (in the original article they use a different 
MST algorithm due to Prim \cite{clrs2001}) and the characteristics of the MSTs is 
studied.

\begin{SCfigure}[][b]
%\centering
\includegraphics[width=0.57\textwidth]{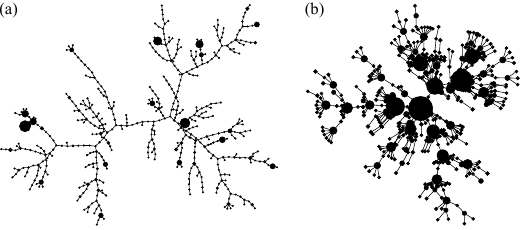}
\caption{
MST of weighted scale free graphs ($N=500$, $m=2$). The size of the nodes 
reflect their degree in $G$.
(a) Disorder type (i): most hubs are located on the outer branches.
(b) Disorder type (ii): hubs of $G$ are at the center of the MST, intermediate
degree nodes (in $G$) are located on the branches of the MST.
\label{fig:exampleGraphs}}
\end{SCfigure}

\begin{figure}
\centering
\includegraphics[width=0.49\textwidth]{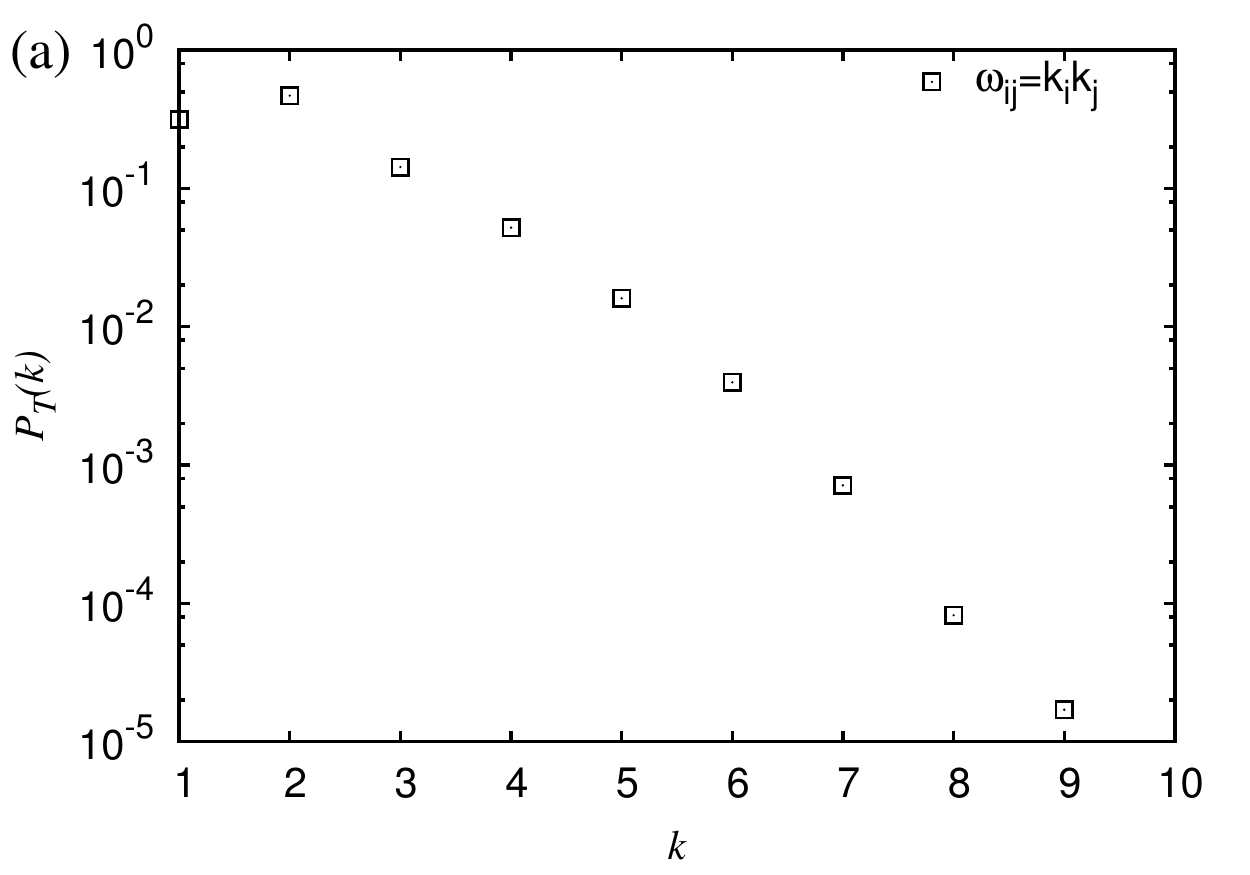}
\includegraphics[width=0.49\textwidth]{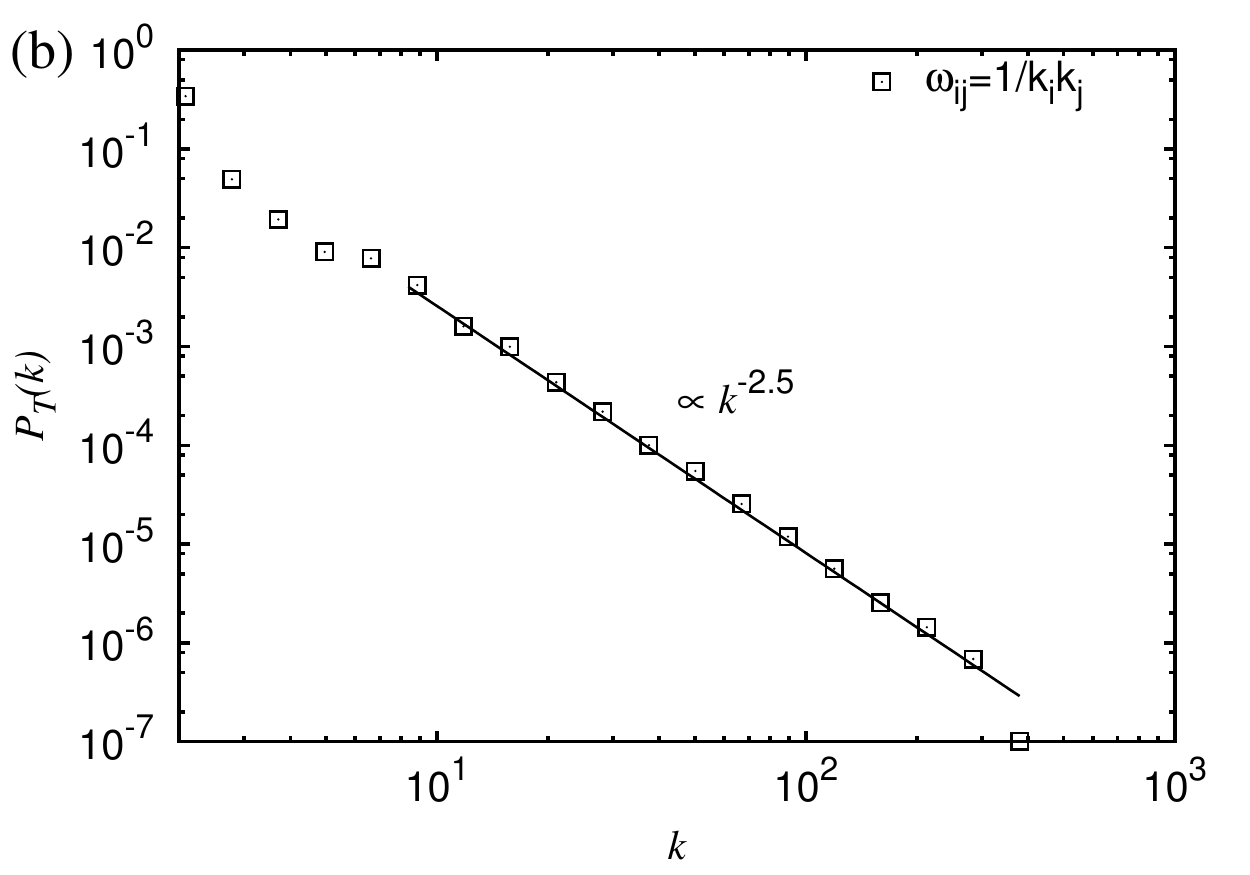}
\includegraphics[width=0.49\textwidth]{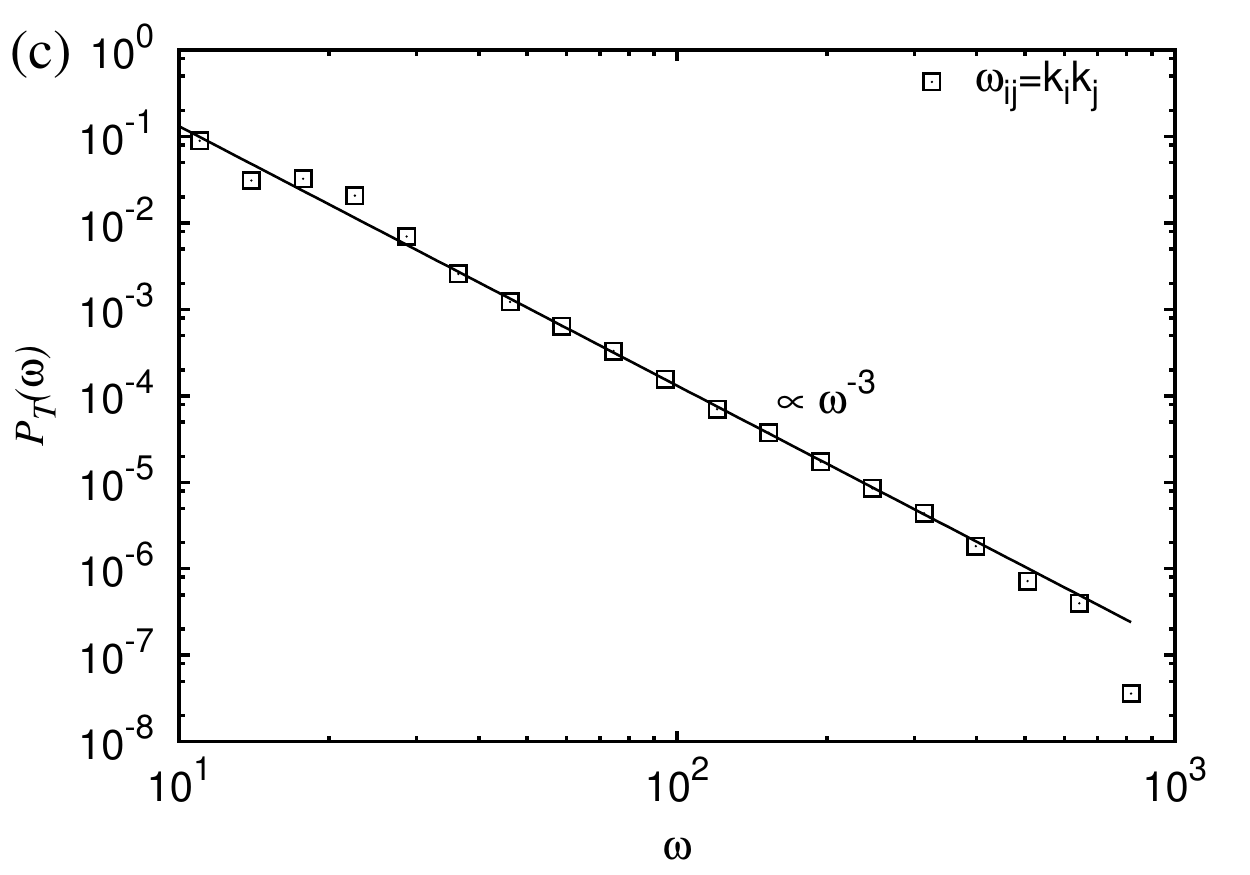}
\includegraphics[width=0.49\textwidth]{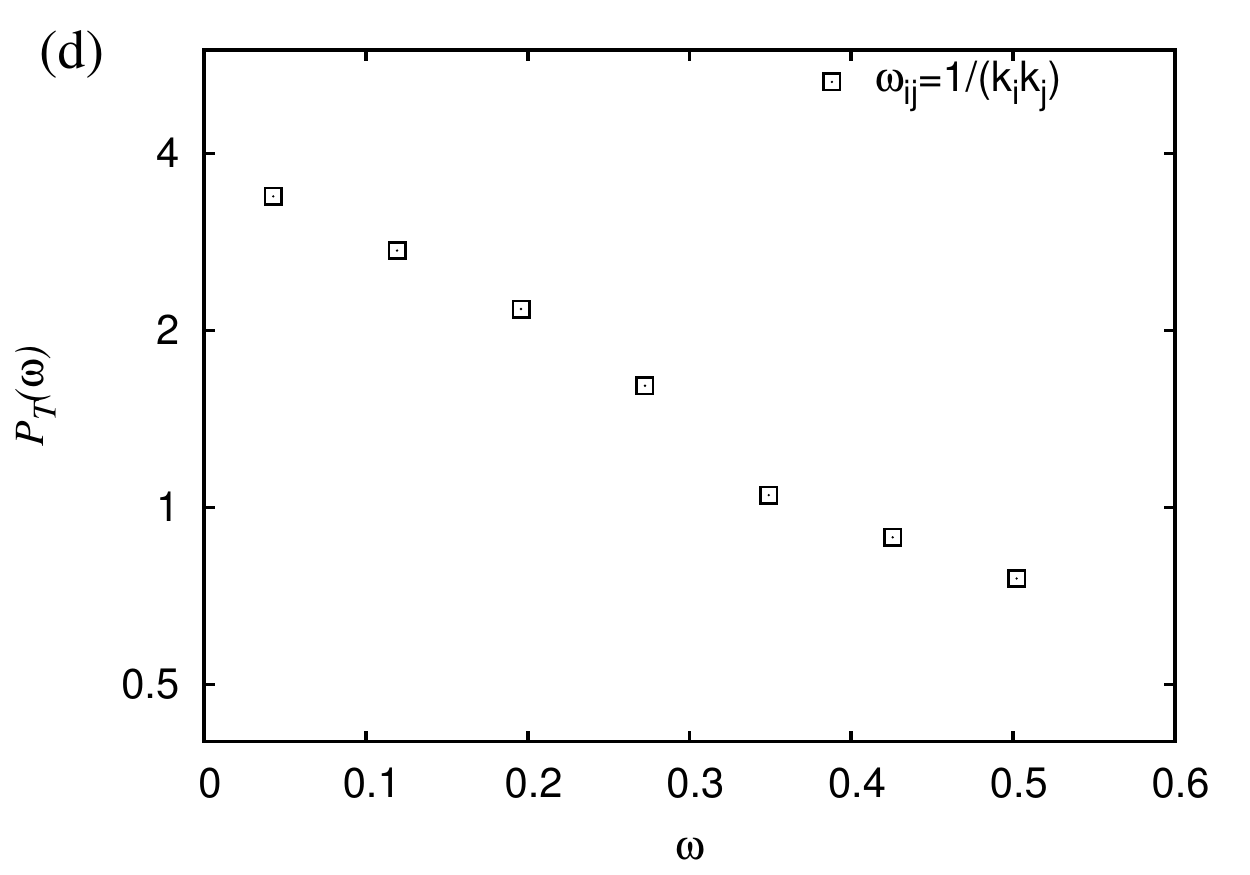}
\includegraphics[width=0.49\textwidth]{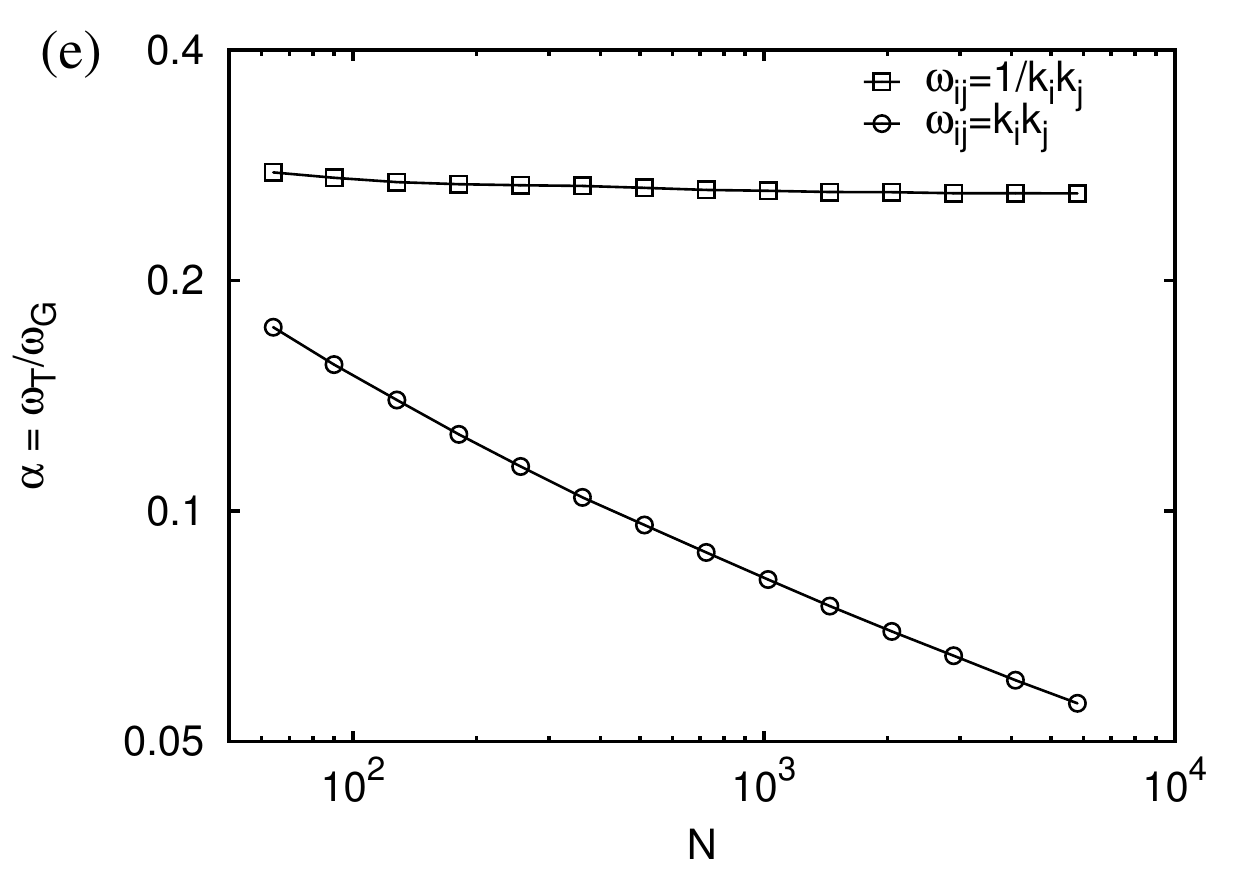}
\caption{
Results for the MSTs computed for weighted scale free graphs ($N=10000$, $m=2$).
(a) degree distribution for weight assignment $\omega_{ij}=k_ik_j$,
(b) degree distribution for weight assignment $\omega_{ij}=1/k_ik_j$,
(c) weight distribution for weight assignment $\omega_{ij}=k_ik_j$,
(d) weight distribution for weight assignment $\omega_{ij}=1/k_ik_j$,
(e) MST efficiency.
\label{fig:results}}
\end{figure}

\paragraph{Results:}
Considering the degree distribution of the nodes regarding the MSTs, the results indicate that
the topology of the MSTs falls into two distinct classes:
\begin{enumerate}
\item Disorder type 1 (Eq.\ (\ref{eq:wgtDistrib1})) yields MST with exponential degree distribution. The MST avoids
edges with large weight, preferentially using edges that connect to low degree nodes (see Figs.\ \ref{fig:results}(a)
 and \ref{fig:exampleGraphs}(a)).
\item Disorder type 2 (Eq.\ (\ref{eq:wgtDistrib2})) yields MST with power law degree distribution. 
Edges with lowest weight are connected to the hubs of $G$. The MST uses these edges extensively 
(see Figs.\ \ref{fig:results}(b) and \ref{fig:exampleGraphs}(b)).
\end{enumerate}

%\begin{figure}

Considering the weight distribution of the edges that comprise the MSTs, the results show that
\begin{enumerate}
\item Disorder type 1 yields MSTs with a power law weight distribution (see Fig.\ \ref{fig:results}(c)).
\item Disorder type 2 yields MSTs with an exponential weight distribution (see Fig.\ \ref{fig:results}(d)).
\end{enumerate}

A measure that tells how efficient a MST connects the nodes of a graph $G$ is given by the 
MST efficiency $\alpha= \omega_T/\omega_G$. Where the numerator signifies the
weight $\omega_T=\sum_{\{i,j\}\in T} \omega_{ij}$ of the MST, and where the denominator
specifies the weight of the graph $\omega_G=\sum_{\{i,j\}\in E} \omega_{ij}$.
The numerical results regarding the MST efficiency are shown in Fig.\ \ref{fig:results}(e).
As evident from the figure, disorder type 1 exhibits a power law decrease of the 
efficiency with increasing graph size.
In contrast to this, the MST efficiency for disorder type 2 saturates at a finite value of $\alpha$.
This suggests that MSTs for disorder type 1 are more efficient than those obtained for disorder type 2.

%-----------------------------------------------------------------------------%
%                          R E F E R E N C E S                                %
%-----------------------------------------------------------------------------%
% For BibTeX users only ... (they will know what to do ...)
%\bibliographystyle{plain}      % we use the plain style
%\bibliography{MCS_LNMelchertRefs}      % expects file "MCS_LNrefs.bib"

\begin{thebibliography}{10}

\bibitem{supMat}
O. Melchert. 
\newblock The supplementary material can be downloaded from the site
  \url{http://www.mcs.uni-oldenburg.de/}.

\bibitem{kruskal1956}
J.~B. Kruskal.
\newblock {\em On the Shortest Spanning Subtree of a Graph and the Traveling Salesman Problem}.
\newblock Proceedings of the American Mathematical Society, 7 (1956) 48. 

\bibitem{clrs2001}
T.~H. Cormen, C.~E. Leiserson, R.~L. Rivest, and C.~Stein.
\newblock {\em {Introduction to Algorithms, 2nd edition}}.
\newblock MIT Press, 2001.

\bibitem{ahuja1993}
R.~K. Ahuja, T.~L. Magnanti, and J.~B. Orlin.
\newblock {\em Network Flows: Theory, Algorithms, and Applications}.
\newblock Prentics Hall, 1993.

\bibitem{gower1969}
J.~C. Gower, and G.~J.~S. Ross.
\newblock {\em Minimum spanning trees and single linkage cluster analysis}.
\newblock Applied Statistics, 18 (1969) 54.

\bibitem{pyRef}
Python is a high-level general-purpose programming language that can be applied
  to many different classes of problems, see \url{http://www.python.org/}.

\bibitem{timsort}
For more information on the Timsort algorithm, see
  \url{http://en.wikipedia.org/wiki/Timsort}.

\bibitem{mcdonald2005}
P.~J. McDonald, E. Almaas, and A.-L. Barab\'asi.
\newblock {\em Minimum spanning trees of weighted scale-free networks}. 
\newblock Europhys. Lett., 72 (2005) 308.

\bibitem{batagelj2005}
V. Batagelj, and U. Brandes.
\newblock {\em Efficient generation of large random networks}.
\newblock Phys.\ Rev.\ E, 71 (2005) 036113.


\end{thebibliography}

\end{document}